\documentclass[aps,prb,reprint,a4paper,floatfix,groupedaddress,amsmath,amssymb,amsfonts]{revtex4-1}
\usepackage{newtxtext,newtxmath}
\usepackage[utf8]{inputenx} 
\usepackage{graphicx}
\usepackage{xspace}
\usepackage[dvipsnames]{xcolor}
\usepackage[pdftex]{hyperref}
\hypersetup{
	pdftitle={Signature of free carriers in Raman spectra of cubic In$_2$O$_3$},
	colorlinks,
	citecolor=blue
	}
\usepackage[
,textwidth=17.5cm
,textheight=23.7cm
,verbose
,dvips
]{geometry} 

\begin{document}

\title{Signatures of free carriers in Raman spectra of cubic In$_2$O$_3$}

\author{Manfred Ramsteiner}
\affiliation{Paul-Drude-Institut f\"ur Festk\"orperelektronik, Leibniz-Institut im Forschungsverbund Berlin e.~V., Hausvogteiplatz 5--7, 10117 Berlin, Germany}
\email{ramsteiner@pdi-berlin.de}
\author{Johannes Feldl}
\affiliation{Paul-Drude-Institut f\"ur Festk\"orperelektronik, Leibniz-Institut im Forschungsverbund Berlin e.~V., Hausvogteiplatz 5--7, 10117 Berlin, Germany}
\author{Zbigniew Galazka}
\affiliation{Leibniz-Institut f\"ur Kristallz\"uchtung, Max-Born-Straße 2, 12489 Berlin, Germany}

\begin{abstract} 
We discuss the influence of free carriers on the Raman scattering in $n$-type In$_2$O$_3$. For high-quality cubic single crystals, electronic single-particle excitations are revealed as a relatively broad Raman feature in the frequency range below 300~cm$^{-1}$. Furthermore, discrete phonon lines in the same frequency range exhibit asymmetric lineshapes characteristic for Fano resonances. The two observed spectral features contain the potential to be utilized for the quantitative determination of the free carrier concentration in $n$-type In$_2$O$_3$ using Raman spectroscopy as a contactless experimental technique.
\end{abstract}

\maketitle

Indium oxide, In$_2$O$_3$, is a transparent semiconducting
oxide which is widely used as transparent contact material in its heavily Sn doped form. The growing interest in advanced applications, for example in transparent logic elements or conductometric gas sensors, is triggered by the unique properties of crystalline In$_2$O$_3$.\citep{bierwagen2015a} For such applications, controlled electrical doping is of crucial importance. Typically, In$_2$O$_3$ exhibits an unintentional $n$-type bulk conductivity in addition to a surface electron accumulation layer and lacks $p$-type conductivity.\citep{bierwagen2015a} Regarding the optical characterization of electrical doping in In$_2$O$_3$, infrared absorption spectroscopy and spectroscopic ellipsometry have been successfully utilized.\citep{fujiwara2005a,galazka2014a,feneberg2016a} Raman spectroscopy has been proven as another powerful method to study free carriers in various materials.\citep{klein1975a,abstreiter1984a} The commonly offered microscopic spatial resolution is one benefit of Raman spectroscopy. The possibility to adjust the optical probing depth via the appropriate choice of the excitation wavelength constitutes a further advantage, in particular regarding the investigation of thin films. At the same time, resonance effects can often be exploited to enhance the sensitivity for the detection of Raman signals.\citep{klein1975a,abstreiter1984a}
However, the commonly analyzed coupled plasmon-phonon modes cannot be observed in Raman spectra of cubic (bixbyite) In$_2$O$_3$. For centrosymmetric bixbyite crystals only even-parity phonon modes are Raman active \citep{garciadomene2012a,kranert2014a} for which the formation of coupled plasmon-phonon is not expected.\citep{klein1975a,abstreiter1984a} Instead, spectral Raman features related to single-particle excitations in free carrier gases as well as uncoupled plasmon modes have been reported for different centrosymmetric semiconductors.\citep{contreras1985a,cerdeira1973a,cerdeira1973b,burke2010a,cerdeira1984a,wagner1985a,mortet2018a}

The aim of the present work is to identify the signatures of free carriers in Raman spectra of $n$-type In$_2$O$_3$ having the potential to be utilized for the evaluation of characteristic quantities such as the free electron concentration. Our study on high-quality single crystals reveals a spectral feature due to a continuum of electronic single-particle excitations as well as an accompanying Fano interference with discrete phonon lines.\citep{fano1961a}

\begin{figure}[b]
\includegraphics*[width=7.0cm]{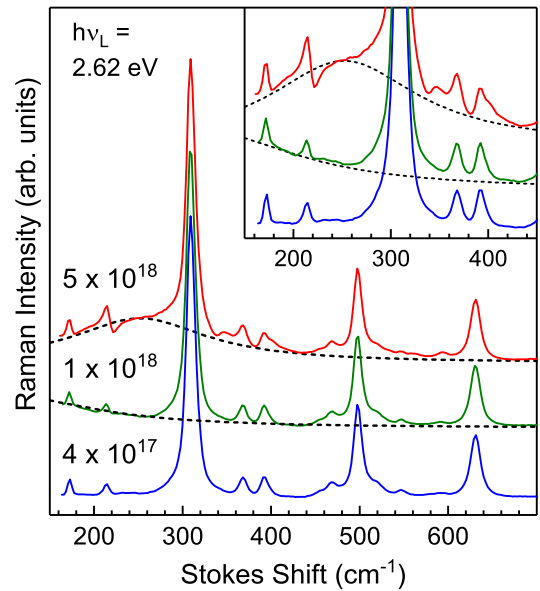}
\caption{\label{fig:1}(Color online) Room-temperature Raman spectra of $n$-type In$_2$O$_3$ crystals with electron concentrations of 5$\times$10$^{18}$ (red), 1$\times$10$^{18}$ (green), and 4$\times$10$^{17}$~cm$^{-3}$ (blue) excited at 2.62~eV. The inset shows the same spectra magnified by a factor of 2. for a smaller frequency range.}
\end{figure}

The investigated cubic bixbyite In$_2$O$_3$ single crystals were 
grown from the melt by the so-called levitation-assisted self-seeding crystal growth.\citep{galazka2013a} In order to achieve different doping levels, the crystals were annealed in different atmospheres leading to free electron concentrations (Hall mobilities) in the range of 4$\times$10$^{17}$ to 5$\times$10$^{18}$~cm$^{-3}$ (180 to 140~cm$^2/$Vs) as determined by Hall effect measurements.\citep{galazka2013a,galazka2014a} Raman measurements were performed in backscattering geometry from (111) surfaces of the In$_2$O$_3$ crystals with the sample temperature controlled by a continuous-flow cryostat. The 405-nm (3.06~eV) and 473-nm (2.62~eV) lines of solid-state lasers and the 632.8-nm (1.96~eV) line of a He-Ne laser were used for optical excitation. The incident laser light was focused by a microscope objective onto the sample surfaces. The backscattered light was collected by the same objective without analysis of its polarization, spectrally dispersed by an 80-cm spectrograph (LabRam HR Evolution, Horiba/Jobin Yvon) and detected with a liquid-nitrogen-cooled charge-coupled device (CCD).

\begin{figure}
\includegraphics*[width=7.0cm]{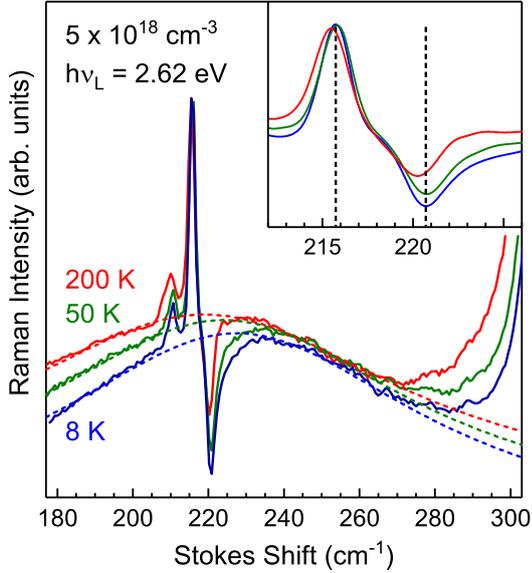}
\caption{\label{fig:2}(Color online) Raman spectra of $n$-type In$_2$O$_3$ with an electron concentration of 5$\times$10$^{18}$~cm$^{-3}$ recorded at temperatures of 200 (red), 50 (green) and 8 K (blue) with excitation at 2.62~eV. The inset shows the same spectra for a smaller frequency range.}
\end{figure}

Figure~\ref{fig:1} displays room-temperature Raman spectra of In$_2$O$_3$ crystals with different carrier densities for excitation at 2.62~eV. The spectra are dominated by the optical phonon lines at 310, 500 and 630~cm$^{-1}$. Additional weaker phonon lines are resolved at 170, 215, 370, and 390~cm$^{-1}$.\citep{kranert2014a,garciadomene2012a} As expected for centrosymmetric crystals, none of the observed phonon lines exhibits a significant plasmon-coupling induced change in intensity or frequency with increasing doping concentration.\citep{klein1975a,abstreiter1984a} In the case of the highest carrier concentration of 5$\times$10$^{18}$~cm$^{-3}$, however, the low-frequency phonon lines are superimposed on a relatively broad Raman band around 250~cm$^{-1}$ (dashed line in Fig.~\ref{fig:1}) and exhibit an asymmetric Fano lineshape.\citep{fano1961a} For the sample with the medium carrier concentration of 1$\times$10$^{18}$~cm$^{-3}$, the contribution of a broad Raman feature at low frequencies is still observable (see dashed line). This kind of spectral features have been observed previously in Raman spectra of different degenerately doped semiconductors.\citep{contreras1985a,cerdeira1973a,cerdeira1973b,burke2010a,cerdeira1984a,wagner1985a,mortet2018a,harima2000a,mitani2012a} The broad band is commonly attributed to electronic excitations in the conduction band.\citep{klein1975a,abstreiter1984a} The Raman peak of such single-particle intraband excitations generally extends over the frequency range $0 < \omega < qv_\mathrm{F}$, where $q$ is the wavevector transferred during the Raman scattering process and $v_\mathrm{F}$ is the Fermi velocity.\citep{voitenko1983a} For the sample with the highest carrier concentration, $qv_\mathrm{F}$ is in the range of 100~cm$^{-1}$ which is clearly below the peak position of the broad band. However, in the case of overdamped excitations, the following relaxation expression is valid to describe the lineshape for electronic intraband Raman scattering:\citep{contreras1985a,zawadowski1990a,ponosov2012a}
\begin{equation}
I_\mathrm{e}(\omega) \propto \frac{1}{1-e^{-\hbar \omega / k_\mathrm{B} T}}\frac{\omega \Gamma}{\omega^2 + \Gamma^2}
\label{spe}
\end{equation} 
where $k_B$ is the Boltzmann constant, $T$ is the sample temperature, and $\Gamma$ is the electron relaxation rate. The resulting peak position of the Raman band coincides roughly with the value of this relaxation rate. In the case of the most heavily doped sample, $\Gamma \approx$ 250~cm$^{-1}$ (see Fig.~\ref{fig:1}) corresponds to a carrier mobility of $\mu_\mathrm{el} = q_\mathrm{e} / m_\mathrm{eff} \Gamma \approx$ 180~cm$^2/$Vs (with the  elementary charge $q_\mathrm{e}$ and the effective mass $m_\mathrm{eff}$), in reasonable agreement with the measured Hall mobility of 140~cm$^2/$Vs.\citep{galazka2013a} However, the fitting of the broad Raman band with Eq.~\ref{spe} does not lead to a satisfactory result. Finally, the lineshape of the broad band could, in principle, be influenced by intersubband transitions between confined states in the surface electron accumulation layer of In$_2$O$_3$.\citep{nagata2017a,nagata2019a,abstreiter1984a} This possibility can be ruled out since the occurence of the broad low-frequency Raman band depends strongly on the bulk carrier concentration (cf. Fig.~1). Consequently, further work has to be done to explain more accurately the lineshape of the observed broad Raman band. Once a satisfactory lineshape model is available, the broad low-frequency Raman band could be used for the analysis of free carrier parameters in In$_2$O$_3$.

\begin{figure}
\includegraphics*[width=7.0cm]{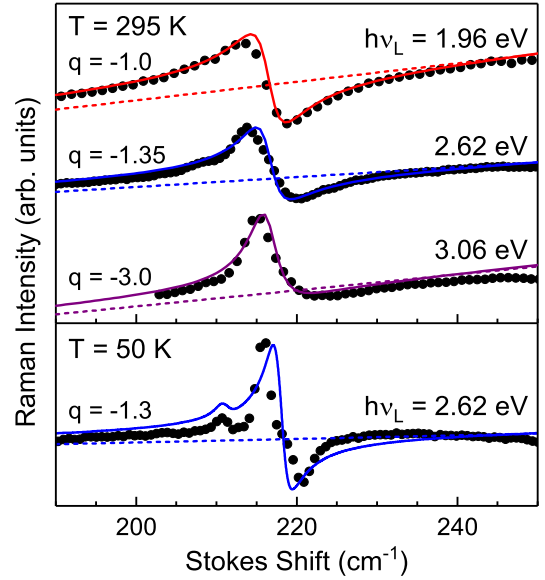}
\caption{\label{fig:3}(Color online) Top: Room-temperature Raman spectra of $n$-type In$_2$O$_3$ with an electron concentration of 5$\times$10$^{18}$~cm$^{-3}$ excited at 1.96 (red), 2.62 (blue), and 3.06~eV. Bottom: Raman spectra of $n$-type In$_2$O$_3$ with an electron concentration of 5$\times$10$^{18}$~cm$^{-3}$ recorded at temperatures of 50~K with excitation at 2.62~eV.}
\end{figure}

The temperature evolution of the Raman spectra in the low-frequency range is shown in Fig.~\ref{fig:2} for the heavily doped sample. With decreasing temperature, the phonon line at 215~cm$^{-1}$ slightly shifts to higher frequencies (see inset of Fig.~\ref{fig:2}). Raman features related to electronic excitations are expected to be paricularly temperature dependent, e.~g., due to changes in the Fermi distribution function and the carrier mobility.\citep{voitenko1983a} Indeed, with decreasing temperature the broad low-frequency Raman feature becomes more narrow and shifts to somewhat higher frequencies (see dashed lines in Fig.~\ref{fig:2}). At the same time, the Fano asymmetry of the phonon line at 215~cm$^{-1}$ becomes more pronounced. These observations support the assignment of the broad Raman band to electronic single-particle excitations. The Fano-type lineshape $I_\mathrm{p}(\omega)$ for discrete phonon scattering interfering these electronic excitations is given by \citep{fano1961a,wagner1985a}
\begin{equation}
I_\mathrm{p}(\omega) = A \frac{(q+\epsilon)^2}{1+\epsilon^2} \\\
\quad\mathrm{with}\quad 
\label{fano1}
\end{equation} 
\begin{equation}
\epsilon = \frac{\omega-\omega_{0}-\Delta\omega}{\gamma}
\quad\mathrm{and}\quad 
q = \frac{\eta R_{ph}}{\gamma R_{el}}.
\label{fano2}
\end{equation} 
Here, $A$ describes the intensity of electronic scattering without phonon interference, $\omega_{0}$ is the frequency of the discrete phonon line in undoped material, $\Delta\omega$ is the renormalization shift due to the electron-phonon interaction, and $\gamma$ is the phonon linewidth. The electron-phonon matrix element is given by $\eta$ and the Raman efficiency for pure phonon scattering by $R_{ph}$. The efficiency for Raman scattering by electronic excitations $R_{el}$ contains information about characteristic free carrier parameters such as the electron concentration.\citep{cerdeira1973b} Assuming the broad Raman band to be described by Eq.~\ref{spe}, the spectra in the low-frequency range are given by the superposition $I_\mathrm{e}(\omega)+ I_\mathrm{p}(\omega)$.

\begin{figure}
\includegraphics*[width=7.0cm]{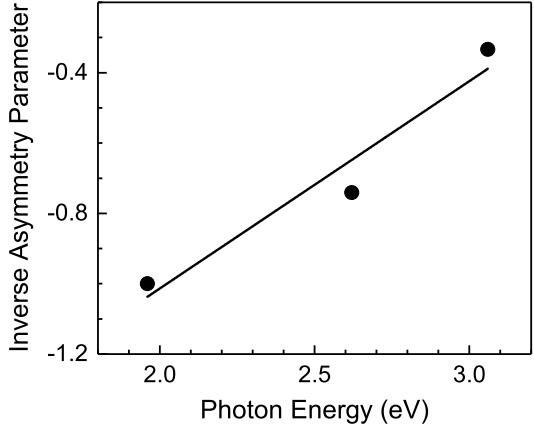}
\caption{\label{fig:4} Inverse asymmetry parameter $q^{-1}$ for $n$-type In$_2$O$_3$ with an electron concentration of 5$\times$10$^{18}$~cm$^{-3}$ as a function of the photon energy used for the excitation of room-temperature Raman spectra.}
\end{figure}

Figure~\ref{fig:3} displays low-frequency Raman spectra (symbols) of the most heavily doped In$_2$O$_3$ sample excited at different photon energies. The agreement with simulations obtained by linshape fittings according to Eqs.~\ref{fano1} and \ref{fano2} (solid lines) confirms the occurence of the Fano-type inferference. For the fitting, the broad Raman band has been taken into account as a linear background signal (see dashed lines in Fig.~\ref{fig:3}).\citep{inushima2003a} The lineshapes of the room-temperature spectra can be explained by a constant phonon energy but different asymmetry parameters $q$. Please note, that the line narrowing at low temperatures reveals the contribution of a second weaker phonon line at the low-frequency side (see Fig.~\ref{fig:2}) leading to a somewhat lower agreement between measured spectrum and fitting curve (see bottom spectrum in Fig.~\ref{fig:3}).

The asymmetry of the Fano-type lineshape resulting from the interaction between phonon scattering and electronic excitations depends on the relative Raman efficiencies of the underlying scattering mechanisms (see Eq.~\ref{fano2}).\citep{cerdeira1973b,cerdeira1973a} Because of the different resonance behaviors of vibrational and electronic Raman scattering, the asymmetry in the Fano lineshape is, in general, expected to depend on the photon energy used for optical excitation. Indeed, the absolute value of the inverse asymmetry parameter $q^{-1}$ obtained for the room-temperature spectra of the most heavily doped In$_2$O$_3$ sample decreases with increasing photon energy as shown in Fig.~\ref{fig:4}. Such a monotonic dependence of $q^{-1}$ on the excitation energy has been found, e.g., for heavily doped silicon when approaching resonance conditions with energy gaps at critical points in the electronic band structure.\citep{cerdeira1973a,burke2010a} In the case of $p$-type silicon, a linear increase of $|q|^{-1}$ has been observed for photon energies in the range between 2.1 and 2.7~eV resulting from the specific resonance behaviors of the involved electronic and vibrational Raman scattering mechanisms.\citep{cerdeira1973a} In the present case, the decrease of $|q|^{-1}$ is attributed to the approaching resonance conditions with the fundamental (2.75~eV) and optical (3.8~eV) band gaps of In$_2$O$_3$.\citep{feneberg2016a} For a fixed excitation condition, however, the magnitude of the asymmetry parameter $q$, as well as the renormalization shift $\Delta\omega$ and the linewidth $\gamma$ (see Eqs.~\ref{fano1} and \ref{fano2}) have been demonstrated to be quantitative measures for the carrier concentration in different materials once calibrated by using an appropriate sample series.\citep{cerdeira1973b,harima2000a,mitani2012a} Furthermore, the lineshape of the Raman peak due to electronic single-particle excitations (see Fig.~\ref{fig:1}) might contain information about the carrier mobility (see Eq.~\ref{spe}). However, further work is needed to determine the dependence of the lineshape parameters (see Eqs.~\ref{spe} to \ref{fano2}) on the doping concentration and mobility. Such a calibration has to be done at a fixed excitation energy (see Fig.~\ref{fig:4}) utilizing a sample series with the carrier concentration varying in a large range.

In conclusion, electronic single-particle excitations and associated Fano interferences with discrete phonon lines are the signatures of free carriers in Raman spectra of $n$-type In$_2$O$_3$. With further calibration work, these spectral features constitute the basis for the evaluation of the free carrier concentration in In$_2$O$_3$ without the need of electrical contacts.

We thank Rüdiger Goldhahn, Martin Feneberg as well as Braulio Garcia Domene for fruitful discussions and Oliver Bierwagen as well as Tobias Schulz for critical reading of our manuscript. This work was performed in the framework of GraFOx, a Leibniz-ScienceCampus partially funded by the Leibniz association. J.F. gratefully acknowledges the financial support by the Leibniz Association.

\end{document}